\documentclass[prb,preprint]{revtex4-1} 
\usepackage{amsmath}  % needed for \tfrac, \bmatrix, etc.
\usepackage{bbm}
\usepackage{amsfonts} % needed for bold Greek, Fraktur, and blackboard bold
\usepackage{graphicx} % needed for figures
\usepackage{verbatim} % needed for figures
 \usepackage{tikz}
\usetikzlibrary{math} 
\usepackage{xcolor}
\begin{document}

% Be sure to use the \title, \author, \affiliation, and \abstract macros
% to format your title page.  Don't use lower-level macros to  manually
% adjust the fonts and centering.

\title{Photon under repeated transverse Lorentz boosts: \\An apparent paradox.}
% In a long title you can use \\ to force a line break at a certain location.

%When submitting the manuscript for review, do not include the author's name or institution
\author{Tugdual LeBohec\footnote{tugdual.lebohec@gmail.com}}
%\email{lebohec@physics.utah.edu} % optional
\affiliation{Department of Physics, University of Utah, Salt Lake City, UT 84112-0830}

%\author{David P. Jackson}
%\email{ajp@dickinson.edu}
%\affiliation{Department of Physics, Dickinson College, Carlisle, PA 17013}

% See the REVTeX documentation for more examples of author and affiliation lists.

\date{\today}

\begin{abstract}
We investigate the effects of the repeated application of Lorentz-boosts to the four momentum of a photon in the transverse direction and observe that this can take us to a reference frame in which the direction of the photon's momentum is apparently reversed. We further extend this to an infinite succession of infinitesimal transverse Lorentz-boosts and show it amounts to a rotation of the photon's momentum, while the transformation is not a simple rotation but a Lorentz transformation. These possibly surprising results  can be understood in light of the Wigner rotation:  the combination of Lorentz-boosts along different directions amounts to a Lorentz-boost combined with a rotation. The presented exercises are some more illustrations of the counterintuitive behavior of Lorentz-boost combinations. \end{abstract}
% AJP requires an abstract for all regular article submissions.
% Abstracts are optional for submissions to the "Notes and Discussions" section.

\maketitle % title page is now complete

%================================================================
%================================================================
\section{Introduction} 
It is commonly known and taught in physics classes that Lorentz-boosts and spatial rotations put together constitute the homogeneous Lorentz group \cite{jackson} of transformations preserving the quadratic form $t^2-x^2-y^2-z^2$ of time $t$ and space coordinates $\{x,y,z\}$. While spatial rotations constitute a group of their own, Lorentz-boosts, mixing time and the coordinate along one direction of space in an hyperbolic analogous to a rotation do not constitute a group when different spacial directions are considered. This asymmetry between the two types of transformations commonly leads to confusion and even misunderstanding \cite{belz}. It seems that even Einstein was puzzled by this, following the discovery of the electronic spin\cite{pais}, which provided an explanation for the fine structure splitting, but with the wrong magnitude until the introduction of what is now known as the Thomas factor\cite{cohentanoudji}. This relativistic correction precisely results from the tracking of the classically circular motion of the electron, which involves compositions of Lorentz-boosts, from which emerges a rotation preventing the simple parallel transport of the electronic spin otherwise expected in classical kinematics \cite{fisher}. What is particularly striking with the Thomas factor is that, being equal to 2, it is not of the order of the speed of the particle in terms of the speed of light as one would expect without realizing what is really going on.  
      
\begin{figure}[h!]
\centering

\begin{tikzpicture}
\tikzmath{\sc=0.95; \sz = \sc*4.0; \st = \sc*4.2; \ph=\sc*0.25; \vel=\sc*1; \hsq3=0.866;}

%K0
\draw[thick,->] (0,0.5*\sz) -- (\sz,0.5*\sz) node[anchor=north east] {$x_0$};
\draw[thick,->] (0.5*\sz ,0) -- (0.5*\sz ,\sz ) node[anchor=north east] {$y_0$};
\draw (0.5*\sz -0.3,0.5*\sz -0.3) node {$K_0$};
\draw[blue,very thick, -stealth] (0.5*\sz, 0.5*\sz)--(0.5*\sz+\ph, 0.5*\sz) node[anchor=south west] {${\bf p}_0$};
\draw[red,very thick, -stealth] (0.5*\sz+0*\st, 0.5*\sz)--(0.5*\sz+0*\st, 0.5*\sz-\vel) node[anchor=south west] {${\bf v}_{1/0}$};

%K1
\draw[thick,->] (0+1*\st,0.5*\sz) -- (\sz+1*\st,0.5*\sz) node[anchor=north east] {$x_1$};
\draw[thick,->] (0.5*\sz+1*\st ,0) -- (0.5*\sz+1*\st ,\sz ) node[anchor=north east] {$y_1$};
\draw (0.5*\sz -0.3+1*\st,0.5*\sz -0.3) node {$K_1$};
\draw[blue,very thick, -stealth] (0.5*\sz+1*\st, 0.5*\sz)--(0.5*\sz+1*\st+0.5*2*\ph, 0.5*\sz+\hsq3*2*\ph) node[anchor=south west] {${\bf p}_1$};
\draw[red,very thick, -stealth] (0.5*\sz+1*\st, 0.5*\sz)--(0.5*\sz+1*\st+\hsq3*\vel, 0.5*\sz-0.5*\vel) node[anchor=north west] {${\bf v}_{2/1}$};

%K2
\draw[thick,->] (0+2*\st,0.5*\sz) -- (\sz+2*\st,0.5*\sz) node[anchor=north east] {$x_2$};
\draw[thick,->] (0.5*\sz+2*\st ,0) -- (0.5*\sz+2*\st ,\sz ) node[anchor=north east] {$y_2$};
\draw (0.5*\sz -0.3+2*\st,0.5*\sz -0.3) node {$K_2$};
\draw[blue,very thick, -stealth] (0.5*\sz+2*\st, 0.5*\sz)--(0.5*\sz+2*\st-0.5*4*\ph, 0.5*\sz+\hsq3*4*\ph)  node[anchor=south east] {${\bf p}_2$};
\draw[red,very thick, -stealth] (0.5*\sz+2*\st, 0.5*\sz)--(0.5*\sz+2*\st+\hsq3*\vel, 0.5*\sz+0.5*\vel)  node[anchor=south west] {${\bf v}_{3/2}$};

%K3
\draw[thick,->] (0+3*\st,0.5*\sz) -- (\sz+3*\st,0.5*\sz) node[anchor=north east] {$x_3$};
\draw[thick,->] (0.5*\sz+3*\st ,0) -- (0.5*\sz+3*\st ,\sz ) node[anchor=north east] {$y_3$};
\draw (0.5*\sz -0.3+3*\st,0.5*\sz -0.3) node {$K_3$};
\draw[blue,very thick, -stealth] (0.5*\sz+3*\st, 0.5*\sz)--(0.5*\sz+3*\st-8*\ph, 0.5*\sz)  node[anchor=south west] {${\bf p}_3$};

\end{tikzpicture}  
\caption{In the laboratory reference frame $K_0$, a photon has a momentum ${\bf p}_0$ along the $\hat{\bf x}_0$-direction. An observer in reference frame $K_1$ in motion with a velocity ${\bf v}_{1/0}$ observes the photon with a momentum ${\bf p}_1$ making an angle $\pi/3$ with the $\hat {\bf x}_1$-direction. We can then consider a second and a third observer in respective reference frames $K_2$ and $K_3$, each in motion transversely to the momentum of the photon in the previous reference frame. Each observer measures the photon momentum in their own reference frame to match calculations by application of a Lorentz-boost to the momentum of the photon observed  in the previous reference frame, each time with a change of direction by an angle $\pi/3$. At each stage, the photon energy is doubled as shown in Section \ref{transvboost}.}
\label{paradox}
\end{figure}
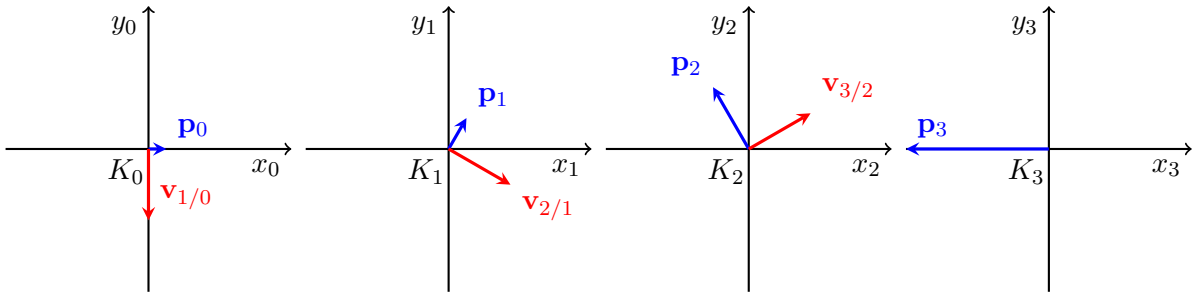

We would like to explore this confusing behavior of combined Lorentz-boosts while taking advantage of the simplification provided by considering a free and massless particle. For this, we are going to apply Lorentz-boosts to the momentum of a photon to reveal an apparent paradox. A Lorentz-boost along a direction that is not collinear with the momentum of the photon results in a reference frame in which the direction  and also, generally, the energy of the photon are changed.  This already suggests the possibility that the combination of several well chosen Lorentz-boosts could  amount to a reversal of the direction of the momentum of the photon. Let's consider the situation presented in Figure \ref{paradox} as an example, which will actually be discussed in details in Section \ref{discrete}. Each observer can measure the photon momentum  in their own reference frame to match what they would calculate by application of a Lorentz-boost to the measurements by the previous observer in the successive reference frames, $K_0$, $K_1$, and $K_2$. Observers in reference frame $K_0$ and $K_3$ could conclude that they observe the same photon with its momentum pointing in two opposite directions. This is interesting as there is no single Lorentz-boost that could relate the respective observations of the same photon in reference frames $K_0$ and $K_3$.  We verify this in Section \ref{LBarbitdir}, where we make use of the expression of a Lorentz-boost in an arbitrary direction obtained in Appendix \ref{arbitboost}. %This apparent paradox results from the fact the combination of several Lorentz boosts is a Lorentz transformation combining a Lorentz boost and a rotation as is shown in Appendix \ref{combinedboosts}. 

In order to clarify the above apparent paradox in details, in Section \ref{transvboost}, we investigate the effect of a Lorentz-boost in a transverse direction on the momentum of a photon.  As we find that the direction of the momentum changes by an angle that depends only on the speed parameter of the transverse Lorentz-boost, in Section \ref{discrete}, we proceed to the $n$-fold repeated application of Lorentz-boosts with the same speed parameter but each time along a direction perpendicular to the momentum of the photon in the current reference frame. We obtain that, as expected, the direction of the momentum changes by an angle that is $n$ times the change in direction resulting from a single transverse Lorentz-boost. This leads to Section \ref{limit}, where we investigate the same process taken to the limit of an infinite number of infinitesimal Lorentz-boosts to obtain an arbitrary change of direction of the photon's momentum without any change in energy. The transformation is however not a simple rotation. Instead it is again a full Lorentz transformation involving both a rotation and a Lorentz-boost, whose velocity parameters (speed and direction) are such that the photon energy is not changed. Finally, section \ref{Summary} is used to summarize our findings and conclude. 

%================================================================
%================================================================
\section{A photon under a Lorentz-boost with an arbitrary direction}\label{LBarbitdir} % Section titles are automatically converted to all-caps.
We wish to establish if a Lorentz-boost can take us to a reference frame in which the direction of the momentum of a photon is reversed. For this, let us consider a photon of energy $\epsilon$ and momentum along the $\hat{\bf x}$-direction in the laboratory reference frame $K_0$. Its energy-momentum four-vector is ${\bf E}_{K_0}\equiv(\epsilon,\epsilon,0,0)$ and we place ourselves in a different reference frame $K_{(\beta,\theta)}$ moving at speed $\beta$ along a direction making an angle $\theta$ with respect to $\hat{\bf x}$  in the $(\hat{\bf x},\hat{\bf y})$ plane, using the Lorentz-boost $\Gamma(\beta,\theta)$ described in Appendix \ref{arbitboost}.  We then see that 
$$\Gamma(\beta,\theta){\bf E}_{K_0}\equiv
\begin{pmatrix}
(1-\beta\cos\theta)\gamma \epsilon\\
  (-\beta\cos\theta+\cos^2\theta+\sin^2\theta/\gamma)\gamma \epsilon\\
\sin\theta((\frac{\gamma-1}{\gamma})\cos\theta-\beta)\gamma \epsilon\\
0 
\end{pmatrix}.
$$
In the new reference frame, the $\hat {\bf y}$-component of the momentum vanishes for $\theta=0$ and $\theta=\pi$, that is when the Lorentz-boost is applied along the photon momentum direction, with it or against it, and, respectively, $\Gamma(\beta,\theta){\bf E}_{K_0}\equiv((1\mp\beta)\gamma \epsilon,  \gamma(1\mp\beta)\gamma \epsilon,0,0 )$. In the new reference frame, $K_{(\beta,0)}$ or $K_{(\beta,\pi)}$, the $\hat {\bf x}$-component of the momentum is respectively smaller or larger than in $K_0$, but it remains positive. In both cases, the momentum of the photon is observed to point in the $+{\bf\hat x}$-direction. 

 For $\sin\theta\ne0$, the $\hat {\bf y}$-component of the momentum trivially vanishes for $\beta=0$, when $\Lambda(\beta,\theta)$ is the identity, or when $\cos\theta=\frac{\beta\gamma}{\gamma -1}$, which happens only in the limit $\beta\to 1$ which places us back in the case  $\theta=0$ (as otherwise $\frac{\beta\gamma}{\gamma -1}>1$). So we see there is no Lorentz-boost taking us from the reference frame $K_0$ in which the photon's momentum points in the ${\bf\hat x}$-direction to a reference frame in which the photon's momentum points in the $-{\bf\hat x}$-direction. 

In the next sections, we will confirm that combining several Lorentz-boosts along different directions can result in a transformation taking us to a reference frame in which the photon momentum is apparently reversed, thus illustrating the fact that a combination of Lorentz-boosts  along different directions does not amount to a Lorentz-boost. 

%================================================================
%================================================================
\section{A photon under transverse Lorentz-boosts }\label{transvboost}
 
Starting again from a photon with the energy-momentum four-vector ${\bf E}_{K_0}\equiv(\epsilon,\epsilon,0,0)$, we apply a Lorentz-boost along a direction perpendicular to the photon's momentum: 
$$\Gamma(\beta,-\pi/2){\bf E}_{K_0}\equiv
\begin{pmatrix}
\gamma \epsilon\\
\epsilon\\
\beta\gamma\epsilon\\
0
\end{pmatrix}.$$ 
In the new reference frame, which we can denote $K_\alpha$,  the photon's momentum makes an angle $\alpha$ with the ${\bf\hat x}$-direction, such that $\tan\alpha=\beta\gamma$ or $\beta=\sin\alpha$, and $\gamma=\frac{1}{\cos\alpha}$, so we have 
$$\Gamma(\sin\alpha,-\pi/2){\bf E}_{K_0}\equiv
\begin{pmatrix}
\epsilon/\cos\alpha\\
\epsilon\\
\epsilon\tan\alpha\\
0
\end{pmatrix}.$$

Consider now the  $n$-fold application of Lorentz-boosts, each one along a direction perpendicular to the direction of the photon momentum obtained from the previous one. All $n$ Lorentz-boosts have the same speed parameter $\beta=\sin\frac{\alpha}{n}$, in such a way the combination takes us to a reference frame $K_{(n,\alpha)}$ in which the photon's momentum makes an angle $n\frac{\alpha}{n}=\alpha$ with the $\bf{\hat x}$-direction, and its energy is $\epsilon/(\cos\frac{\alpha}{n})^n$. It should be noted that $K_\alpha=K_{(1,\alpha)}$ but for $n>1$, $K_\alpha\ne K_{(n,\alpha)}$ as, while the angle between the photon momentum and the ${\bf\hat x}$-direction are the same in both, the energy is different. 

To take a specific example, we could choose $\alpha=\pi$ and $n=3$, in which case, each Lorentz-boost is applied in a direction perpendicular to the photon momentum given by the previous one with $\beta=\sin\frac{\pi}{3}=\frac{\sqrt 3}{2}$ and $\gamma=2$. After $n=3$ such Lorentz-boosts, the four-momentum of the photon in the final reference frame $K_{(3,\pi)}$ would be $(8\epsilon,-8\epsilon,0,0)$. This is precisely what is illustrated in Figure \ref{paradox}. As we saw in Section \ref{LBarbitdir}, there is no single Lorentz-boost reverting the direction of the momentum of the photon. This may seem paradoxical if one does not remember that the combination of several Lorentz-boosts along different directions does not amount to a simple Lorentz-boost but to the combination of a Lorentz-boost with a rotation \cite{ferraro1999} as is shown in Appendix \ref{combinedboosts}. The rotation emerging from Lorentz-boosts composition is known as Wigner's rotation \cite{wigner1939}. The $n$-fold combined transformation is studied in detail in the next Section \ref{discrete}.   Pushing this to the limit $n\to\infty$ in the general case of an angle $\alpha$, we can expect the four-momentum of the photon in the resulting reference frame would tend to $(\epsilon,\epsilon\cos\alpha,\epsilon\sin\alpha,0)$, because $\lim_{n\to\infty}\left[\cos(\alpha/n)\right]^n$=1. This limit transformation will be studied in Section \ref{limit}. 

%================================================================
%================================================================
\section{A photon under repeated transverse Lorentz-boosts}\label{discrete}

In the previous section, we understood the effect of repeated transverse Lorentz-boosts on the momentum of a photon but we did not obtain an expression for the combined transformation and, in particular, we do not know the relative velocities of the final reference frame $K_{(n,\alpha)}$ and the laboratory reference frame $K_0$ with respect to one-another, which we are now going to establish.

The Lorentz-boost along the $-\hat{\bf y}$-direction is $\Gamma_y(-\beta)=\Gamma(\beta,-\pi/2)$ and, with a photon of energy $\epsilon$ and momentum pointing along the $\hat{\bf x}$-direction in the laboratory reference frame $K_0$, we have already seen that $\Gamma_y(-\sin\alpha)$ takes us to a reference frame where the photon has an energy $\epsilon/\cos\alpha$ and a momentum making an angle $\alpha$ with the $\hat{\bf x}$-direction.

In order to build a transformation $\Lambda(n,\alpha)$ amounting to $n$ successive applications of $\beta=-\sin(\alpha)$ Lorentz-boosts, each along a direction perpendicular to the momentum of the photon after the previous Lorentz-boost, we can proceed similarly as in Appendix \ref{arbitboost}, using the rotation $R_{\bf{\hat z}}(\alpha)$ in combination with the Lorentz-boost $\Gamma_y(-\sin\alpha)$ so as to track the photon momentum direction as follows, where, in order to lighten notation we use $\Gamma_\alpha=\Gamma_y(-\sin\alpha)$ and $R_\alpha=R_{\bf{\hat z}}(\alpha)$, with the property  $\left( R_\alpha\right)^k=R_{k\alpha}$ extending to the reciprocals with $k<0$:
\begin{align}
\Lambda(1,\alpha)&=\Gamma_\alpha \nonumber\\
\Lambda(2,\alpha)&=R_\alpha\Gamma_\alpha R_{-\alpha}\Gamma_\alpha\nonumber\\
\Lambda(3,\alpha)&=R_{2\alpha}\Gamma_\alpha R_{-2\alpha}R_\alpha\Gamma_\alpha R_{-\alpha}\Gamma_\alpha\nonumber\\
&\cdots&\nonumber\\
\Lambda(n,\alpha)&=R_{(n-1)\alpha}\Gamma_\alpha\left[R_{-\alpha}\Gamma_\alpha\right]^{n-1} \label{combination}
\end{align}
Using a symbolic calculator\cite{mathematica2023}, we verify that $\Lambda(n,\alpha)$ takes us from the laboratory reference frame $K_0$, where a photon of energy $\epsilon$ and momentum pointing along the $\hat{\bf x}$-direction to a reference frame $K_{(n,\alpha)}$, in which the same photon has an energy $\frac{\epsilon}{(\cos\alpha)^n}$ and a momentum pointing along a direction making an angle $n\alpha$ with the $\hat{\bf x}$-direction, just as expected already from the discussion in Section \ref{transvboost}. In particular, $\Lambda(n,\pi/n)$ brings the photon along the $-\hat{\bf x}$-direction for any $n>2$. 

\begin{figure}[h!]
\centering
\includegraphics[width=11cm]{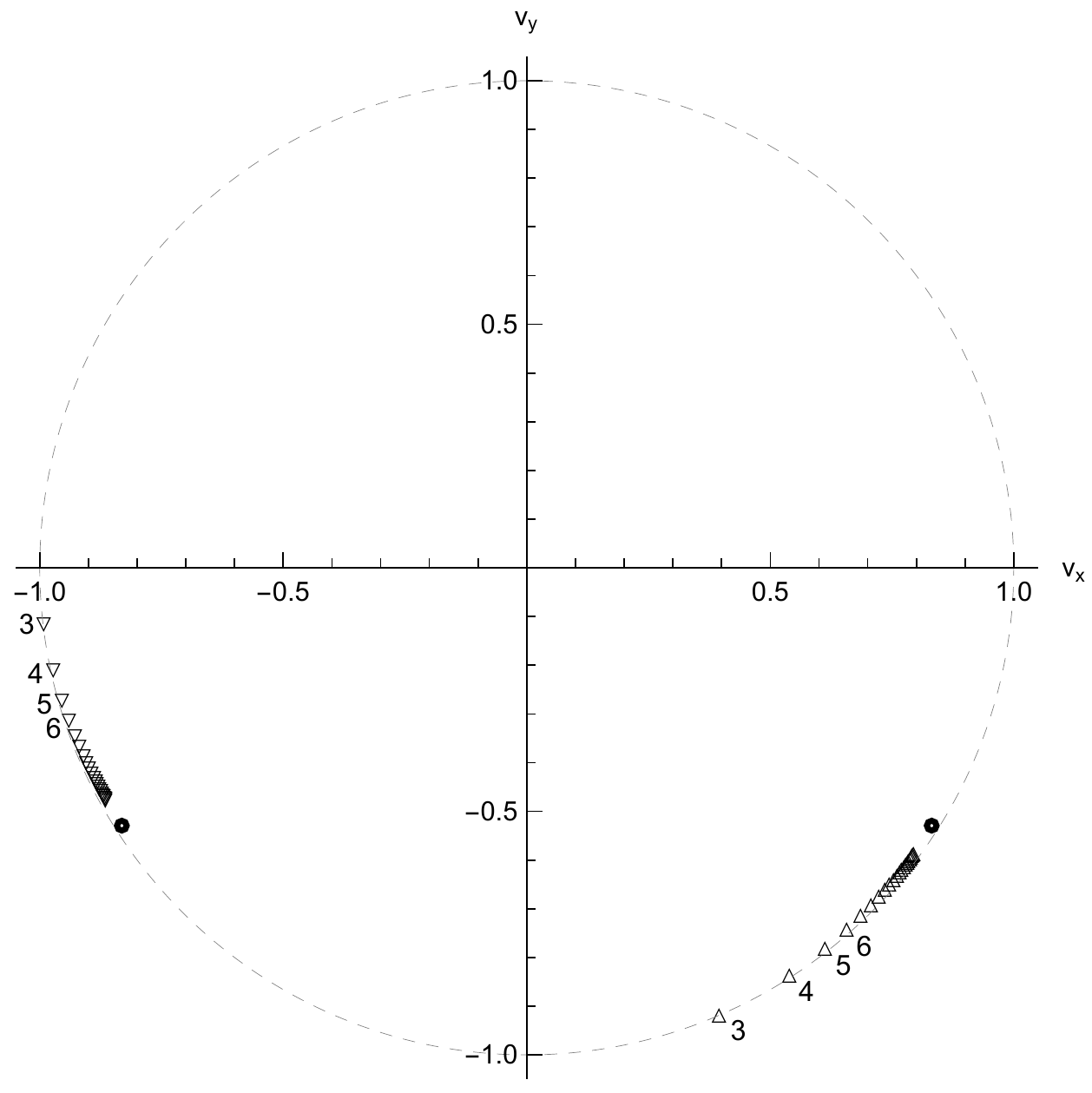}
\caption{The mutually relative velocities of the laboratory $K_{0}$ and $K_{(n,\alpha)}$ reference frames for the special case $\alpha=\pi$ with $n$ values ranging from $3$ to $24$. Pointing down triangles represent the velocities ${\bf v}^{(0/\pi)}_n$ of the laboratory $K_{0}$ reference frame with respect to the $K_{(n,\pi)}$ reference, frame while pointing down triangles represent the velocities ${\bf v}^{(\pi/0)}_n$ of the $K_{(n,\pi)}$ reference frame with respect to the laboratory reference frame $K_0$. The values of $n$ from 3 to 6 are indicated next to the corresponding points in both cases. For small values of $n$, the points representing ${\bf v}_n^{(0/\pi)}$ and ${\bf v}_n^{(\pi/0)}$ are very close to the unit circle and they slowly depart from it as $n$ increases. The small circles indicate the $n\to\infty$ limits of respective velocities ${\bf v}_\infty^{(0/\pi)}$ and ${\bf v}_\infty^{(\pi/0)}$ obtained in Section \ref{limit}.\label{relatvel}}
\label{gasbulbdata}
\end{figure}

We can obtain ${\bf v}^{(0/n,\alpha)}_n$, the velocity of the laboratory reference frame $K_0$ with respect to the $K_{(n,\alpha)}$ reference frame, by acting with $\Lambda(n,\alpha)$ on the four-velocity of the laboratory reference frame with respect to itself, ${\bf V}^{(0/0)}\equiv(1,0,0,0)$. Inversely, we can obtain ${\bf v}^{(n,\alpha/0)}_n$, the velocity of the $K_{(n,\alpha)}$ reference frame with respect to the laboratory reference frame, by acting on ${\bf V}^{(0/0)}$ with the reciprocal transformation $\Lambda^{-1}(n,\alpha)$, which can be obtained by taking advantage of the fact $\Gamma^{-1}_\alpha=\Gamma_{-\alpha}$ as
\begin{equation}
\Lambda^{-1}(n,\alpha)=\left[\Gamma_{-\alpha}R_{\alpha}\right]^{n-1}\Gamma_{-\alpha} R_{-(n-1)\alpha}.\label{reversecombination}
\end{equation}
We find that for $n>1$ and $\alpha\ne0$, as a result of the Wigner rotation emerging from the $n$-fold Lorentz-boost composition (See Appendix \ref{combinedboosts}), and differently from what is familiar with non-relativistic changes of reference frames, or changes of reference frames via a single Lorentz-boost (Appendix \ref{arbitboost}), ${\bf v}^{(0/n,\alpha)}_n\ne-{\bf v}^{(n,\alpha/0)}_n$. This is illustrated in Figure \ref{relatvel} for the special case $\alpha=\pi/n$ with a range of values of $n>2$ so as to avoid $\beta=1$. It appears that, as we approach the limit $n\to\infty$, for which we expect the energy of the photon to be the same $\epsilon$ as in the laboratory, the velocities ${\bf v}^{(0/\alpha)}_\infty$ and ${\bf v}^{(\alpha/0)}_\infty$ maintain some angle between them. We investigate this limit in the next section.

%================================================================
%================================================================
\section{The limit $n\to\infty$}\label{limit}

In order to obtain $\Lambda(n,\alpha/n)$ in the limit $n\to\infty$, we can proceed as above, using $\Gamma_y(\beta)$ and $R_{\bf{\hat z}}(\beta)$ in the limit $\beta\to0$, where we have $\beta=\alpha/n$.
Equation \ref{combination} brings our attention to  $R_{\bf{\hat z}}(-\beta)\Gamma_{\bf{\hat y}}(\beta)$ and, in the limit $\beta\ll 1$, 
$$
R_{\bf{\hat z}}(-\beta)\Gamma_{\bf{\hat y}}(\beta)\equiv
		\begin{pmatrix}
		1 &      0 &       0 & 0\\
		0 &      1 &\beta & 0\\
		0 &-\beta &       1 & 0\\
		0 &      0 &       0 & 1 
		\end{pmatrix}	\cdot	
		\begin{pmatrix}
		      1 & 0 & \beta & 0\\
		      0 & 1 &      0  & 0\\
		\beta & 0 &       1 & 0\\
		      0 & 0 &       0 & 1 
		\end{pmatrix}=
		\begin{pmatrix}
		                 1 &            0 & \beta& 0\\
		       \beta^2 &            1 &\beta & 0\\
		           \beta &    -\beta &       1 & 0\\
		                 0 &           0 &       0 & 1 
		\end{pmatrix}.
$$
With this, anticipating the limit evaluation, we may drop out the quadratic term and write $R_{\bf{\hat z}}(-\beta)\Lambda_{\bf{\hat y}}(\beta)=\mathbbm{1}+\beta\tau_{\bf{\hat {zy}}}$
with
$$
\tau_{\bf{\hat {zy}}}\equiv
		\begin{pmatrix}
		       0 &           0 &       1& 0\\
		       0 &           0 &       1 & 0\\
		       1 &          -1 &       0 & 0\\
		       0 &           0 &       0 & 0 
		\end{pmatrix}.
$$
Similarly, we write $R_{\bf{\hat z}}(\beta)=\mathbbm{1}+\beta\rho_{\bf{\hat z}}$ with $\rho_{\bf{\hat z}}$
the generator of the $R_{\bf{\hat z}}(\beta)$ rotations. Then, using $\rho_{\bf{\hat z}}$ and $\tau_{\bf{\hat {zy}}}$ in Equation \ref{combination} in the limit $\beta\ll 1$ and $n\gg 1$, we write  
$$
\Lambda(n,\beta)=\left(\mathbbm{1}+\beta\rho_{\bf{\hat z}}\right)^{n-1}\Lambda_{\bf{\hat y}}(\beta)\left(\mathbbm{1}+\beta\tau_{\bf{\hat {zy}}}\right)^{n-1}.
$$
We can then choose $\beta=\alpha/(n-1)$ and take the limit $n\to\infty$ to define $\Lambda(\alpha)=e^{\alpha\rho_{\bf{\hat z}}}e^{\alpha\tau_{\bf{\hat {zy}}}}$,
where we used the fact that $\lim_{\beta\to 0}\Lambda_{\bf{\hat y}}(\beta)=\mathbbm{1}$.  

We already know that $e^{\alpha\rho_{\bf{\hat z}}}=R_{\hat{\bf z}}(\alpha)$ and we need to identify $e^{\alpha\tau_{\bf{\hat {zy}}}}$.  For this, we hope for the best and calculate 
$$
\tau_{\bf{\hat {zy}}}^2\equiv
%		\begin{pmatrix}
%		       0 &           0 &       -1& 0\\
%		       0 &           0 &      -1 & 0\\
%		      -1 &           1 &       0 & 0\\
%		       0 &           0 &       0 & 0 
%		\end{pmatrix}\cdot
%		\begin{pmatrix}
%		       0 &           0 &       -1& 0\\
%		       0 &           0 &      -1 & 0\\
%		      -1 &           1 &       0 & 0\\
%		       0 &           0 &       0 & 0 
%		\end{pmatrix}=
		\begin{pmatrix}
		       1 &          -1 &       0& 0\\
		       1 &          -1 &       0 & 0\\
		       0 &           0 &       0 & 0\\
		       0 &           0 &       0 & 0 
		\end{pmatrix},
$$
and note with relief that $\tau_{\bf{\hat {zy}}}^3=0$ and so $e^{\alpha\tau_{\bf{\hat {zy}}}}=\mathbbm{1}+\alpha\tau_{\bf{\hat {zy}}}+\frac 1 2 \alpha^2\tau_{\bf{\hat {zy}}}^2$.
%$$
%e^{\alpha\tau}=		
%		\begin{pmatrix}
%		              1+\alpha^2/2 &  -\alpha^2/2 & -\alpha& 0\\
%		                  \alpha^2/2 &1-\alpha^2/2 & -\alpha& 0\\
%		                       -\alpha &        \alpha &         1 & 0\\
%		                               0 &               0 &         0 & 1 
%		\end{pmatrix}.
%$$
Then, combining with the rotation $e^{\alpha\rho_{\bf{\hat z}}}$, we have 

$$
\Lambda(\alpha)=
e^{\alpha\rho_{\bf{\hat z}}}e^{\alpha\tau_{\bf{\hat {zy}}}}\equiv
%		\begin{pmatrix}
%		       1 &               0 &              0 & 0\\
%		       0 & \cos\alpha & \sin\alpha & 0\\
%		       0 &-\sin\alpha & \cos\alpha & 0\\
%		       0 &              0 &              0 & 1 
%		\end{pmatrix}\cdot
%		\begin{pmatrix}
%		              1+\alpha^2/2 &  -\alpha^2/2 & -\alpha& 0\\
%		                  \alpha^2/2 &1-\alpha^2/2 & -\alpha& 0\\
%		                       -\alpha &        \alpha &         1 & 0\\
%		                               0 &               0 &         0 & 1 
%		\end{pmatrix}=
		\begin{pmatrix}
				1+\alpha^2/2                                                &  -\alpha^2/2                                                       &  \alpha                                & 0\\
				\frac \alpha 2(\alpha\cos\alpha-2\sin\alpha)   & (1-\frac{\alpha^2} 2)\cos\alpha+\alpha\sin\alpha & \alpha\cos\alpha-\sin\alpha  & 0\\
				\frac \alpha 2(2\cos\alpha+\alpha\sin\alpha) & (1-\frac{\alpha^2} 2)\sin\alpha-\alpha\cos\alpha & \cos\alpha+\alpha\sin\alpha & 0\\	
				0                                                                  &   0                                                                      &   0                                        & 1			
		\end{pmatrix}.
$$

We can now go back to the original problem starting from a photon with a momentum pointing along the $\bf{\hat x}$-direction with an energy $\epsilon$, and calculate 
$$\Lambda(\alpha)\cdot\begin{pmatrix}\epsilon\\\epsilon\\0\\0\end{pmatrix}\equiv\begin{pmatrix}\epsilon\\\epsilon\cos\alpha\\\epsilon\sin\alpha\\0\end{pmatrix}.$$ We see that, as expected, $\Lambda(\alpha)$ takes us to the reference frame $K_\alpha=\lim_{n\to\infty}K_{n,\alpha/n}$, in which the photon has the same energy $\epsilon$ but a momentum along a direction making an angle $\alpha$ with the $\hat{\bf x}$-direction. In particular, for the case $\alpha=\pi$, just as we anticipated in Section \ref{transvboost}, the photon's energy-momentum in $K_\pi$ is $(\epsilon,-\epsilon,0,0)$. The transformation $\Lambda(\alpha)$  is however not a simple rotation and actually includes a Lorentz-boost. This can be verified by noting that the first column of the matrix representing $\Lambda(\alpha)$ is the four-velocity of the laboratory reference frame with respect to $K_\alpha$, and, in the case $\alpha=\pi$, the corresponding velocity is ${\bf v}_\infty^{(0/\alpha)}=-\frac{\pi^2}{2+\pi^2}\hat{\bf x}-\frac{2\pi}{2+\pi^2}\hat{\bf y}$, as indicated on Figure \ref{relatvel}. 

Starting from Equation \ref{reversecombination}, and following the same procedure, we can obtain the reciprocal $\Lambda^{-1}(\alpha)$ and the matrix representing it, whose first column gives the four-velocity of the $K_\alpha$ reference frame with respect to the $K_0$ reference frame. In the case $\alpha=\pi$, the corresponding velocity is ${\bf v}_\infty^{(\alpha/0)}=\frac{\pi^2}{2+\pi^2}\hat{\bf x}-\frac{2\pi}{2+\pi^2}\hat{\bf y}$, also as indicated on Figure \ref{relatvel}. 

The fact that the photon energy is the same in reference frames $K_0$ and $K_\alpha$ suggests some condition must be satisfied by the mutual relative velocities of these reference frames. Indeed, acting with $\Gamma(\beta,\theta)$ from Appendix \ref{arbitboost} on ${\bf E}_{K_0}$, we see that, for the photon energy to be unchanged, the condition $\gamma(1-\beta\cos\theta)=1$ must be satisfied, and we verify that the condition is satisfied by both ${\bf v}_\infty^{(0/\alpha)}$ and ${\bf v}_\infty^{(0/\alpha)}$ when identifying $\beta$ to their magnitude and  $\theta$ to the angle they make with the momentum of the photon. This is expected as, up to a rotation, single Lorentz-boosts can take us from $K_0$ to $K_\alpha$ and inversely.   

%================================================================
%================================================================
\section{Summary and conclusions}\label{Summary}
We started by considering a Lorentz-boost applied transversally to the momentum of a photon. The operation changes the direction of the photon momentum by an angle $\alpha$ depending only on the Lorentz-boost parameter $\beta$. This drew us toward iteratively combining Lorentz-boosts, each along a direction perpendicular to the momentum of the photon after the previous iteration. With an appropriate choice of the Lorentz-boosts parameter and number of iterations, this can take us to a reference frame in which the photon's momentum is apparently reversed compared to the situation in the laboratory reference frame we started from. We identified that this results from the fact that a combination of Lorentz-boosts in different directions does not amount to a simple Lorentz-boost but to the combination of a Lorentz-boost with a rotation by an angle known as the Wigner angle as shown in Appendix \ref{combinedboosts}. This was investigated using combinations of rotations with the Lorentz-boosts constructed in Appendix \ref{arbitboost} in the case $\theta=-\pi/2$ for a finite number $n$ of iterations, each changing the direction of the photon by an angle $\alpha/n$ in Sections \ref{discrete}, where we obtained $\Lambda(n,\alpha/n)$ and for an infinite number of infinitesimal Lorentz-boosts in Section \ref{limit}, where we obtained $\Lambda(\alpha)$, with which the energy of the photon is not changed. In both cases, we verified the photon momentum direction is changed by an angle $\alpha$. We also observed that the mutually relative velocities of the laboratory $K_0$ and final $K_{(n,\alpha)}$ or $K_\alpha$ reference frames  are not opposite to each other but form an angle, which we identify to the combined Wigner angle emerging from the many Lorentz-boosts combination. 

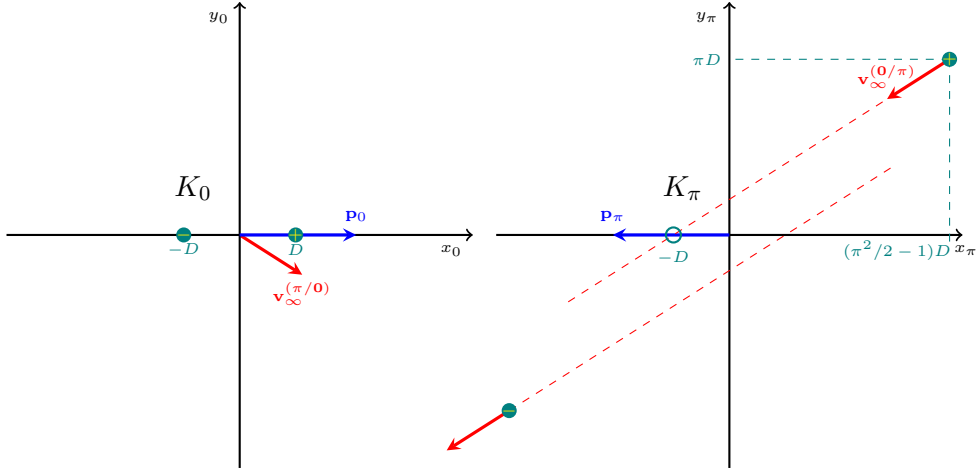
\begin{figure}[h!]
\centering

\begin{tikzpicture}
\tikzmath{\sc=1.55; \sz = \sc*4.0; \st = \sc*4.2; \ph=\sc*1.0;  \dst=0.12;}

%K0
%Kalpha reference frame velocity
\draw[red,very thick, -stealth] (0.5*\sz, 0.5*\sz)--(0.5*\sz+0.8315, 0.5*\sz-0.5293) node[anchor=south east] {};
\draw [red](0.5*\sz+0.8315, 0.5*\sz-0.5293-0.25) node {\tiny$\bf{v}_\infty^{(\pi/0)}$};
%X axis
\draw[thick,->] (0,0.5*\sz) -- (\sz,0.5*\sz) node[anchor=north east] {\tiny$x_0$};
%Y axis
\draw[thick,->] (0.5*\sz ,0) -- (0.5*\sz ,\sz ) node[anchor=north east] {\tiny$y_0$};
%Photon momentum
\draw[blue,very thick, -stealth] (0.5*\sz, 0.5*\sz)--(0.5*\sz+\ph, 0.5*\sz) node[anchor=south west] {};
\draw [blue](0.5*\sz+\ph, 0.5*\sz+0.25) node {\tiny ${\bf p}_0$};
%Frame name
\draw (0.5*\sz -0.1*\sz,0.5*\sz+0.1*\sz ) node {$K_0$};
%Detector
\path [draw=none,fill=teal, fill opacity = 1] (0.5*\sz+\dst*\sz,0.5*\sz) circle (0.1);
\draw [yellow](0.5*\sz +\dst*\sz,0.5*\sz ) node {\tiny$+$};
\draw [teal](0.5*\sz +\dst*\sz,0.5*\sz -0.2) node {\tiny$D$};
\path [draw=none,fill=teal, fill opacity = 1] (0.5*\sz-\dst*\sz,0.5*\sz) circle (0.1);
\draw [yellow](0.5*\sz -\dst*\sz,0.5*\sz ) node {\tiny$-$};
\draw [teal](0.5*\sz -\dst*\sz,0.5*\sz -0.2) node {\tiny$-D$};

%Kpi
%X axis
\draw[thick,->] (0+1*\st,0.5*\sz) -- (\sz+1*\st,0.5*\sz) node[anchor=north east] {};
\draw (\sz+0.05+1*\st,0.5*\sz-0.2) node {\tiny $x_\pi$};
%Y axis
\draw[thick,->] (0.5*\sz+1*\st ,0) -- (0.5*\sz+1*\st ,\sz ) node[anchor=north east] {\tiny$y_\pi$};
%Photon momentum
\draw[blue,very thick, -stealth] (0.5*\sz+1*\st, 0.5*\sz)--(0.5*\sz+1*\st-\ph, 0.5*\sz) node[anchor=south east] {};
\draw [blue](0.5*\sz-\ph+1*\st, 0.5*\sz+0.25) node {\tiny ${\bf p}_\pi$};
%Frame name
\draw (0.5*\sz -0.1*\sz+1*\st,0.5*\sz +0.1*\sz) node {$K_\pi$};
%Detector velocity
\draw[red,very thick, -stealth] (0.5*\sz+3.9348*\dst*\sz+1*\st, 0.5*\sz+3.1416*\dst*\sz)--(0.5*\sz+3.9348*\dst*\sz-0.8315+1*\st, 0.5*\sz+3.1416*\dst*\sz-0.5293) node[anchor=south east] {};
\draw[red,very thick, -stealth] (0.5*\sz-3.9348*\dst*\sz+1*\st, 0.5*\sz-3.1416*\dst*\sz)--(0.5*\sz-3.9348*\dst*\sz-0.8315+1*\st, 0.5*\sz-3.1416*\dst*\sz-0.5293) node[anchor=south east] {};
\draw [red](0.5*\sz+3.9348*\dst*\sz-0.8315+1*\st, 0.5*\sz+3.1416*\dst*\sz-0.5293+0.3) node {\tiny$\bf{v}_\infty^{(0/\pi)}$};
\draw[red,thin,dashed] (0.5*\sz+3.9348*\dst*\sz+1*\st, 0.5*\sz+3.1416*\dst*\sz)--(0.5*\sz+3.9348*\dst*\sz-4*0.8315*\sc+1*\st, 0.5*\sz+3.1416*\dst*\sz-4*0.5293*\sc) node[anchor=south east] {};
\draw[red,thin,dashed] (0.5*\sz-3.9348*\dst*\sz+1*\st, 0.5*\sz-3.1416*\dst*\sz)--(0.5*\sz-3.9348*\dst*\sz+4*0.8315*\sc+1*\st, 0.5*\sz-3.1416*\dst*\sz+4*0.5293*\sc) node[anchor=south east] {};%Detector
\path [draw=none,fill=teal, fill opacity = 1] (0.5*\sz+3.9348*\dst*\sz+1*\st,0.5*\sz+3.1416*\dst*\sz) circle (0.1);
\draw[teal,thin,dashed] (0.5*\sz+3.9348*\dst*\sz+1*\st, 0.5*\sz+3.1416*\dst*\sz)--(0.5*\sz+3.9348*\dst*\sz+1*\st, 0.5*\sz-0.1) node[anchor=south east] {};
\draw[teal,thin,dashed] (0.5*\sz+3.9348*\dst*\sz+1*\st, 0.5*\sz+3.1416*\dst*\sz)--(0.5*\sz+1*\st, 0.5*\sz+3.1416*\dst*\sz) node[anchor=south east] {};
\draw [yellow](0.5*\sz+3.9348*\dst*\sz+1*\st,0.5*\sz+3.1416*\dst*\sz) node {\tiny$+$};

\path [draw=none,fill=teal, fill opacity = 1] (0.5*\sz-3.9348*\dst*\sz+1*\st,0.5*\sz-3.1416*\dst*\sz) circle (0.1);
\draw [yellow](0.5*\sz-3.9348*\dst*\sz+1*\st,0.5*\sz-3.1416*\dst*\sz) node {\tiny$-$};

\path [draw=teal,fill=none, fill opacity = 1,thick] (0.5*\sz-1*\dst*\sz+1*\st,0.5*\sz) circle (0.1);
\draw [teal](0.5*\sz -1*\dst*\sz+1*\st,0.5*\sz -0.3) node {\tiny$-D$};
\draw [teal](0.5*\sz +3.9348*\dst*\sz+1*\st-0.7,0.5*\sz -0.2) node {\tiny$(\pi^2/2-1)D$};
\draw [teal](0.5*\sz -0.3+1*\st,0.5*\sz+3.1416*\dst*\sz ) node {\tiny$\pi D$};
\end{tikzpicture}  
\caption{ Left panel: In the laboratory reference frame $K_0$, two detectors (solid circles labeled $+$ and $-$) are on the $\hat{\bf x}_0$-axis at a distance $D$ from the origin, where a photon is emitted in the $+\hat{\bf x}_0$-direction at time $t=0$ with a momentum ${\bf p}_0$ (blue arrow). The photon will be received by the detector in $x_0=+D$. The velocity $\bf{v}_\infty^{(\pi/0)}$ of the reference frame $K_\alpha$ defined in Section \ref{limit} with $\alpha=\pi$ is also indicated (red arrow). Right panel: In the reference frame $K_\alpha$, the positions of the detectors are obtained by using the Lorentz transformation $\Lambda(\pi)$. The detector labeled $+$ is represented at time $t=-\frac \pi 2 D$ and the one labeled $-$ at time $t=\frac \pi 2 D$. Both are moving along the red dashed lines with velocity $\bf{v}_\infty^{(0/\pi)}$. The Wigner rotation appears in the non-collinearity of $\bf{v}_\infty^{(\pi/0)}$ and $\bf{v}_\infty^{(0/\pi)}$. The photon emitted at time $t=0$ by the origin in the $-\hat{\bf x}_\pi$-direction with a momentum ${\bf p}_\pi=-{\bf p}_0$ and is detected at time $t=+D$, when the detector labeled $+$ crosses the $\hat{\bf x}_\pi$-axis  in $x_\pi=-D$ (open circle). }
\label{detector}
\end{figure}

The apparent reversal of the direction of the photon momentum obtained for $\alpha=\pi$ should not change the sequence of events taking place in a setup including detectors, such as the one described in Figure \ref{detector}. In the laboratory reference frame $K_0$, where the photon is emitted at time $t_0=0$ in the $+\hat{\bf x}_0$-direction, two detectors are on the $\hat{\bf x}_0$-axis, one, labeled $+$, in $x_0=+D$ will collect the photon at time $t_0=D$, and the other, labeled $-$, in $x_0=-D$. The transformation $\Lambda(\pi)$ from Section \ref{limit} takes us to reference frame $K_\pi$ in motion with respect to the laboratory with the velocity ${\bf v}_\infty^{(\pi/0)}$. The events space-time coordinates $(0,\pm D,0,0)$ corresponding to detectors labeled $\pm$ at time $t_0=0$ in $K_0$, are transformed to $(\mp\frac{\pi^2}{2}D,\pm(\frac{\pi^2}{2}-1)D,\pm\pi D,0)$ in $K_\pi$. Both detectors are moving with velocity ${\bf v}_\infty^{(0/\pi)}$ with respect to $K_\pi$. The detector labeled $+$ crosses the $\hat{\bf x}_\pi$-axis in $x_\pi=-D$, just in time ($t_\pi=D$) to collect the photon emitted in $t_\pi=0$. So, we see with relief that the same detector collects the photon both in $K_0$ and $K_\alpha$. The fact the time between photon emission and detection in $K_\pi$ is the same as in $K_0$ corresponds to the fact the space-time interval between emission and detection events is light-like and along the $\hat{\bf x}$-direction in $K_0$. It transforms under $\Lambda(\pi)$, in the same way as the photon energy-momentum, with the time component unchanged and the space component along $\hat{\bf x}$ reversed.  We could construct the same figure for $\Lambda(n, \pi)$ from Section \ref{discrete}. The time between emission and detection would then be changed by the same factor $\left(\cos\frac\pi n\right)^{-n}$ changing the photon energy. 

In these exercises, we focused  on the special case $\alpha=\pi/n$ for $\Lambda(n,\alpha)$ and $\alpha=\pi$ for $\Lambda(\alpha)$, which both result in a photon with a momentum pointing in a direction that is reversed compared to what it is in the laboratory reference frame. Alternatively, we could consider the cases $\alpha=2k\pi$ for $\Lambda(\alpha)$ where $k$ is an integer, which leaves the photon energy-momentum unchanged while the relative speed of the reference frame $K_{(2k\pi)}$ is then $\frac{k\pi}{2+k^2\pi^2}\sqrt{4+k^2\pi^2}$, which tends to the speed of light as $k\to\infty$. 

The calculation of the limit $n\to\infty$ in Section \ref{limit} can be reminiscent of the calculation of the Thomas precession for a particle following a circular orbit. The latter proceeds by the construction of an infinitesimal Lorentz transformation connecting the rest frames of a particle at two different times separated by an infinitesimal interval while the particle follows a circular trajectory\cite{fisher,jackson}. The generator appearing in the infinitesimal transformation is then decomposed into the generators of a Lorentz-boost and a rotation, the Wigner rotation. There are however important differences between the two calculations. Mathematically first, the transformation $\Lambda(\alpha)=e^{\alpha\rho_{\bf{\hat z}}}e^{\alpha\tau_{\bf{\hat {zy}}}}$ is written in terms of  $\rho_{\bf{\hat z}}$, the generator of rotations around $\hat{\bf z}$, and $\tau_{\bf{\hat {zy}}}$ the sum of the generators of rotations around $\hat{\bf z}$ and Lorentz-boosts along $\hat{\bf y}$. As $\rho_{\bf{\hat z}}$ and $\tau_{\bf{\hat {zy}}}$ do not commute\cite{jackson}, $\Lambda(\alpha)$ cannot be written in terms of a single generator\cite{eugene}, which would otherwise allow us to identify a rotation component and a Lorentz-boost component as is done in the calculation of the Thomas precession. Then, the situation is physically quite different. In the calculation of the Thomas precession, the Lorentz transformations are constructed to track in time a physical process, an infinitesimal element of circular motion. In the case of our photon, its energy momentum four vector is constant  as it does not participate in any physical process. We performed exercises in which we considered this same photon from different reference frames, by monitoring the relative directions of the momentum. These exercises are illustrations that Lorentz-boosts, apparently simple linear transformations, can, in combinations, lead to complex, counterintuitive, and entertaining results.

\appendix
%================================================================
%================================================================
\section{Lorentz-boost in an arbitrary direction}\label{arbitboost}
In order to obtain a Lorentz-boost  transformation from the laboratory reference frame $K^{(0)}$ to  reference frame $K^{(1)}$ moving at speed $\beta\in[0,1]$ along a direction making an angle $\theta$ with the $ {\bf\hat x}$-direction, we can first proceed to a rotation by an angle $-\theta$ to a new coordinate system with its $\bf{\hat x}$-axis along the direction of the desired Lorentz-boost, then apply the familiar Lorentz-boost of speed parameter $\beta$ along the $\bf{\hat x}$-direction of the new coordinate system, and finally apply a rotation by an angle $\theta$, reverse to the first one, so as to ensure that, in the reference frame $K^{(1)}$, the reference frame $K^{(0)}$ is in motion along a direction making an angle $\pi-\theta$ with respect to the $\bf{\hat x}$-direction.  This amounts to calculating $\Gamma(\beta,\theta)=R_{\bf{\hat z}}(\theta)\Gamma(\beta,0) R_{\bf{\hat z}}(-\theta)$, where $R_{\bf{\hat z}}(\theta)$ is the rotation by an angle $\theta$ around the $\bf{\hat z}$-direction, 
$$
R_{\bf{\hat z}}(\theta)\equiv
		\begin{pmatrix}
		1 &            0 &           0 & 0\\
		0 & \cos\theta & -\sin\theta  & 0\\
		0 & \sin\theta & \cos\theta & 0\\
		0 &             0 &          0 & 1 \\
		\end{pmatrix},
$$
and $\Gamma(\beta,0)$ is the Lorentz-boost along the $\bf{\hat x}$-direction,
$$
\Gamma(\beta,0)\equiv		\begin{pmatrix}
		\gamma          & -\beta\gamma & 0 & 0\\
		-\beta\gamma &          \gamma & 0  & 0\\
		0                     &                     0 & 1 & 0 \\
		0                     &                     0 & 0 & 1 \\
		\end{pmatrix}.
$$ 
We find
$$
\Gamma(\beta,\theta)\equiv
		\begin{pmatrix}
		\gamma                      & -\beta\gamma\cos\theta            &-\beta\gamma\sin\theta              & 0\\
		-\beta\gamma\cos\theta  & \gamma\cos^2\theta+\sin^2\theta&\cos\theta\sin\theta(\gamma-1) & 0\\
		-\beta\gamma\sin\theta & \cos\theta\sin\theta(\gamma-1) &\cos^2\theta+\gamma\sin^2\theta   & 0\\
		0                                 &                                             0 &                                           0 & 1 \\
		\end{pmatrix}.
$$
It should be noted that, while in the case $\theta=0$ with $\Gamma(\beta,0)$, the images in reference frame $K^{(1)}$ of the $\bf{\hat x}$- and $\bf{\hat y}$-axis of $K^{(0)}$ are mutually perpendicular and in fact coincide with the $\bf{\hat x}$- and $\bf{\hat y}$-axis of $K^{(1)}$, in the case $\theta\ne0$ (and $\theta\ne\pi$), there is a distortion resulting from Lorentz length-contraction \cite{moller52}. The images of 
$\bf{\hat x}$- and $\bf{\hat y}$-axis of $K^{(0)}$ are not mutually perpendicular and do not coincide with the $\bf{\hat x}$- and $\bf{\hat y}$-axis of $K^{(1)}$.

We can now act with $\Gamma(\beta,\theta)$ on the four velocity of the laboratory with respect to the laboratory ${\bf V}^{(0/0)}\equiv(1,0,0,0)$ to obtain ${\bf V}^{(0/1)}=\Gamma(\beta,\theta).{\bf V}^{(0/0)}\equiv(\gamma, -\beta\gamma\cos\theta,-\beta\gamma\sin\theta,0)$ as expected. We also verify that $\Gamma(\beta,\theta)\cdot\Gamma(-\beta,\theta)=\Gamma(\beta,\theta+\pi)\cdot\Gamma(\beta,\theta)=\mathbbm{1}$ and this implies that ${\bf V}^{(1/0)}=\Gamma(-\beta,\theta).{\bf V}^{(1/1)}\equiv(\gamma, \beta\gamma\cos\theta,\beta\gamma\sin\theta,0)$, where we used ${\bf V}^{(1/1)}\equiv(1,0,0,0)$. We see that ${\bf V}^{(0/1)}$ and ${\bf V}^{(1/0)}$ can be obtained from each other by changing the signs of their respective spacial components, as expected. 

\section{Combination of two Lorentz-boosts}\label{combinedboosts}
We can combine two Lorentz-boosts. The first one, $\Gamma(\beta',0)$, takes us from the laboratory reference frame $K^{(0)}$ to another, $K^{(1)}$, moving at speed $\beta'$ along the ${\bf\hat x}$-direction. The second, $\Gamma(\beta,\theta)$ , takes us from $K^{(1)}$ to reference frame $K^{(2)}$ moving at speed $\beta$ along a direction making an angle $\theta$ with the ${\bf\hat x}$-direction. We define $\Lambda_{02}(\beta,\theta,\beta')=\Gamma(\beta,\theta)\Gamma(\beta',0)$, and, with $\gamma'=\frac{1}{\sqrt{1-\beta'^2}}$, we find 
$${\scriptscriptstyle \Lambda_{02}(\beta,\theta,\beta')\equiv}
\left(
\begin{array}{cccc}
{\scriptscriptstyle \gamma'\gamma(1+\beta'\beta\cos\theta)} & {\scriptscriptstyle-\gamma'\gamma(\beta'+\beta\cos\theta)} & {\scriptscriptstyle-\gamma\beta\sin\theta)} & {\scriptscriptstyle0} \\
{\scriptscriptstyle -\gamma'\gamma(\beta\cos\theta+\beta'\cos^2\theta-\beta'\sin^2\theta/\gamma)}& {\scriptscriptstyle\gamma'\gamma(\beta'\beta\cos\theta+\cos^2\theta+\sin^2\theta/\gamma)} & {\scriptscriptstyle(\gamma-1)\sin\theta\cos\theta} & {\scriptscriptstyle0} \\
 {\scriptscriptstyle -\gamma'\gamma (\beta+\beta'(1-1/\gamma)\cos\theta)\sin\theta} & {\scriptscriptstyle \gamma'\gamma(\beta'\beta+\cos\theta(1-1/\gamma))\sin\theta } & {\scriptscriptstyle \cos^2\theta+\gamma\sin^2\theta} & {\scriptscriptstyle 0} \\
 {\scriptscriptstyle0} & {\scriptscriptstyle0} & {\scriptscriptstyle0} & {\scriptscriptstyle1} \\
\end{array}
\scriptscriptstyle\right).
$$

%$${ \Lambda_{02}(\beta,\theta,\beta')\equiv}
%\left(
%\begin{array}{cccc}
%{ \gamma'\gamma(1+\beta'\beta\cos\theta)} & {-\gamma'\gamma(\beta'+\beta\cos\theta)} & {-\gamma\beta\sin\theta)} & {0} \\
%{ -\gamma'\gamma(\beta\cos\theta+\beta'\cos^2\theta-\beta'\sin^2\theta/\gamma)}& {\gamma'\gamma(\beta'\beta\cos\theta+\cos^2\theta+\sin^2\theta/\gamma)} & {(\gamma-1)\sin\theta\cos\theta} & {0} \\
%{ -\gamma'\gamma (\beta+\beta'(1-1/\gamma)\cos\theta)\sin\theta} & { \gamma'\gamma(\beta'\beta+\cos\theta(1-1/\gamma))\sin\theta } & { \cos^2\theta+\gamma\sin^2\theta} & { 0} \\
% {0} & {0} & {0} & {1} \\
%\end{array}
%\right).
%$$

With this we can consider the velocity of the laboratory reference frame $K_0$ with respect to reference frame $K^{(2)}$ by expressing ${\bf V}^{(0/2)}=\Lambda_{02}(\beta,\theta,\beta'){\bf V}^{(0/0)}$ which just gives the first column of the matrix representing $\Lambda_{02}(\beta,\theta,\beta')$, in which we recognize as the standard configuration velocity composition law. 

In order to express ${\bf V}^{(2/0)}$ we could act on ${\bf V}^{(2/2)}$ with the reverse transformation or, instead, we can start from ${\bf V}^{(2/1)}\equiv(\gamma,\gamma\beta\cos\theta,\gamma\beta\sin\theta,0)$ and  calculate ${\bf V}^{(2/0)}=\Gamma(-\beta',0){\bf V}^{(2/1)}$, which gives 
$${\bf V}^{(2/0)}\equiv\begin{pmatrix}\gamma'\gamma(1+\beta'\beta\cos\theta)\\\gamma'\gamma(\beta'+\beta\cos\theta)\\\gamma\beta\sin\theta\\0\end{pmatrix}.$$ 
We see that ${\bf V}^{(0/2)}$ and ${\bf V}^{(2/0)}$ have the same time component and, correspondingly, we verify that the magnitudes, respectively $|{\bf v}^{(0/2)}|$ and $|{\bf v}^{(2/0)}|$, of their space parts also are equal 
$$
|{\bf v}^{(0/2)}|=|{\bf v}^{(2/0)}|=\frac{\sqrt{\beta^2+2\beta\beta'\cos\theta+\beta'^2(1-\beta^2\sin^2\theta)}}{1+\beta\beta'\cos\theta}.  
$$ 
This implies that ${\bf v}^{(0/2)}$ and ${\bf v}^{(2/0)}$ are equal to each other up to a rotation by an angle $\phi$ around direction $\hat{\bf z}$, and we have 
$$
\sin\phi=\frac{v^{(0/2)}_xv^{(2/0)}_y-v^{(0/2)}_yv^{(2/0)}_x}{|{\bf v}^{(0/2)}|\cdot|{\bf v}^{(2/0)}|},
$$
or
$$
\sin\phi=\sin\theta\frac{(\beta\cos\theta+\beta')(\beta+\beta'\cos\theta(1-1/\gamma))-\beta(\beta\cos\theta+\beta'(\cos^2\theta+\sin^2\theta/\gamma))/\gamma}{\beta^2+2\beta\beta'\cos\theta+\beta'^2(1-\beta^2\sin^2\theta)},
$$
while classically we would expect ${\bf v}^{(0/2)}$ and ${\bf v}^{(2/0)}$ to point in opposite directions with $\phi=\pi$.

\begin{figure}[h]
\centering
\includegraphics[width=11cm]{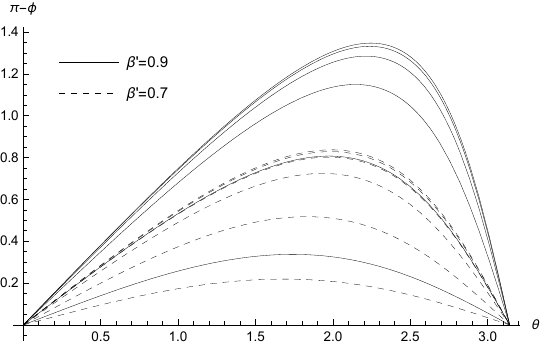}
\caption{\label{wignerrot} The angle $\pi-\phi$ is shown as a function of $\theta$ for $\beta'=0.7$, and $\beta'=0.9$, and $\beta$ with values 0.5, 0.9, 0.99, 0.999, 0.9999, and 0.99999. }.
\end{figure}

This illustrates the fact the composition of Lorentz-boosts along different directions amounts to a Lorentz transformation combining a Lorentz-boost with a rotation. Figure \ref{wignerrot} shows $\pi-\phi$ as a function of $\theta$ for selected values of $\beta$ and $\beta'$. We see that it is only when the Lorentz-boosts are collinear  ($\theta=0$ or $\theta=\pi$), that $\phi=\pi$. In all other cases,  ${\bf v}^{(0/2)}$ and ${\bf v}^{(2/0)}$ do not point in opposite directions. 

We can then finally go back from $K_2$ to the laboratory reference frame $K_0$ by using one more Lorentz-boost to define $\Lambda_{00}(\beta,\theta,\beta')=\Gamma(|{\bf v}^{(0/2)}|,\theta^{(0/2)})\Lambda_{02}(\beta,\theta,\beta')$, where $\theta^{0/2}$ is the angle between ${\bf v}^{(0/2)}$ and the $\hat{\bf x}$-direction. It is impractical and uninteresting to reproduce the matrix representing $\Lambda_{00}$ here but we verify that, as expected, it is a rotation matrix by an angle $\phi'$. Also as expected, when the first two Lorentz-boosts are collinear  ($\theta=0$ or $\theta=\pi$), we verify the third Lorentz-boost also is along the same direction and  $\phi'=0$. 

\begin{acknowledgments}
The author wishes to acknowledge John Belz and Eugene Mishchenko for their helpful and motivating input. 
All the calculations presented in this article were done using Mathematica \cite{mathematica2023}. 

\end{acknowledgments}

\end{document}